\documentclass[aps,preprintnumbers,prl,twocolumn,superscriptaddress]{revtex4}

\usepackage{epsfig,latexsym,cancel,amssymb,amsmath}
\usepackage{graphicx}

\DeclareMathOperator{\erf}{Erf}

\newcommand{\keV}{\ensuremath{\mathrm{keV}}}
\newcommand{\MeV}{\ensuremath{\mathrm{MeV}}}
\newcommand{\GeV}{\ensuremath{\mathrm{GeV}}}
\newcommand{\TeV}{\ensuremath{\mathrm{TeV}}}

\newcommand{\cm}{\ensuremath{\mathrm{cm}}}

\newcommand{\xe}{\ensuremath{\mathrm{Xe}}}
\newcommand{\cs}{\ensuremath{\mathrm{Cs}}}
\newcommand{\xex}{\ensuremath{{}^{136}\mathrm{Xe}}}

\begin{document}
\DeclareGraphicsExtensions{.jpg,.pdf,.mps,.png,}

\title{Direct constraints on charged excitations of dark
  matter}

 \author{Haipeng An}
\affiliation{Perimeter Institute, Waterloo, Ontario N2L 2Y5, Canada}

\author{Maxim Pospelov}
\affiliation{Perimeter Institute, Waterloo, Ontario N2L 2Y5, Canada}
\affiliation{Department of Physics and Astronomy, University of Victoria, Victoria, BC, V8P 5C2 Canada}

\author{Josef Pradler}
\affiliation{Department of Physics and Astronomy, Johns Hopkins University, Baltimore, MD 21210, USA}

\begin{abstract}
  If the neutral component of weak-scale dark matter 
  is accompanied by a charged excitation separated by a mass gap of
  less than $\sim$20~MeV, WIMPs can form stable bound states with
  nuclei. We show that the recent progress in experiments searching
  for neutrinoless double-beta decay sets the first direct constraint
  on the 
  exoergic reaction of WIMP-nucleus bound state formation. We
  calculate the rate for such process in representative models and
  show that the double-beta decay experiments provide unique
  sensitivity to a large fraction of parameter space of the WIMP
  doublet model, complementary to constraints imposed by cosmology and
  direct collider searches.
\end{abstract}

\maketitle

\paragraph{Introduction}
\label{sec:introduction}

Weakly
interacting massive particles (WIMPs) are well-motivated candidates
for dark matter (DM), that offer a variety of potential
non-gravitational signatures, including the possibility for a
laboratory detection.  Model realizations of WIMP DM often entail
multiple states.
Frequently encountered are scenarios in which DM ($X^0$) is part of a
multiplet with an electrically charged excited state ($X^{\pm}$) that
enables to naturally regulate its abundance through co-annihilation.
As has been noted in~\cite{Pospelov:2008qx}, if the mass gap between
the neutral and charged state is sufficiently small,
\textit{MeV-scale} energy depositions become possible via the process
of WIMP-nucleus bound state formation.

\begin{figure}[b]
 \begin{center}
 \begin{tabular}{cc}
   \includegraphics[scale=0.6]{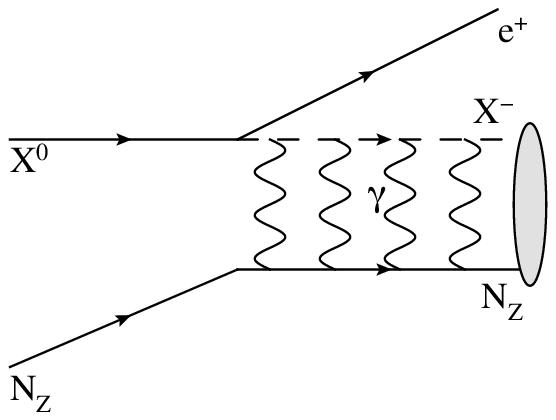}\;\;~&~\;\;\includegraphics[scale=0.6]{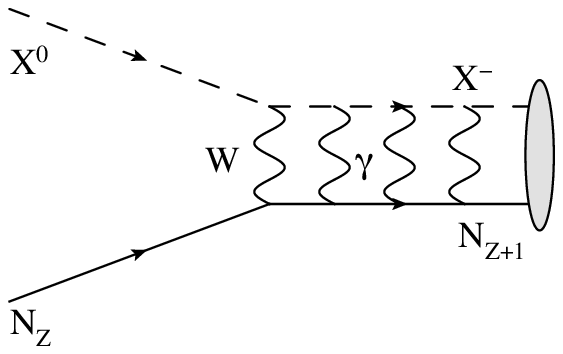} \\
   (a)~&~(b) \\
 \end{tabular} 
 \caption{Illustrations of case A (a) and case B (b).}
 \label{fig:n02beta}
 \end{center}
 \end{figure}

Denote by $(NX^{-})$ the bound state of DM with a target nucleus $N$.
Depending on the relation between spins of $X^0$ and $X^-$ two generic
scenarios can be envisaged,
\begin{align} 
\label{eq:A}
 \text{Case A:}\quad &  N_Z + X^0 \to (N_ZX^{-}) + e^{+}  , \\
\label{eq:B}
 \text{Case B:}\quad &  N_Z + X^0 \to (N_{Z+1}X^{-})  \ , 
\end{align}
where $Z$ denotes the charge of $N$. The Feynman diagrams of these two scenarios are shown in Fig.~\ref{fig:n02beta}.
If $(N X^-)$ is not in its ground state, it will de-excite by emitting
$\gamma$-rays. Thus, the observables for these processes are the
positron and $\gamma$-rays (and a residual anomalously heavy nucleus.)
The recombination is guaranteed to happen once the $(NX^{-})$ Coulomb
binding energy $E_b$ allows to bridge the mass gap $\Delta m \equiv
m_{X^{-}} - m_{X^0}$ in the DM multiplet.
Mass splittings of $\sim$20~MeV or less can be probed through
(\ref{eq:A}) and (\ref{eq:B}).
Case B can arise in the scalar DM models, multiplets of the SM weak group 
\cite{Cirelli:2005uq}, with $\sim$20\% cancellation between tree-level and loop-induced $\Delta m$. 
Both scenarios are readily realized in supersymmetric (SUSY) scenarios.

Recent FERMI-LAT observations~\cite{Fermi} of a 135~GeV photon excess
from the galactic center stimulate studies of WIMP models with mass
splitting on the order of few MeV and less, as a source of enhancement
of WIMP annihilation to mono-energetic
$\gamma$-rays~\cite{Pospelov:2008qx}.
Additional interest to models with $O(\rm MeV)$ mass splitting comes from the 
possibility of DM catalysis of primordial nuclear reactions, changing the outcome for lithium isotopes,
that currently shows unexpected deviations \cite{Pospelov:2010hj}.

The primary purpose of this letter is to show that the recent advances
in neutrinoless double-beta ($0\nu\beta\beta$) decay experiments allow
for a direct probe the neutral-charged WIMP doublet models.
The $O($1-10 MeV) scale energy deposition in (\ref{eq:A}) and
(\ref{eq:B}) give $0\nu\beta\beta$ searches a clear advantage over
direct DM detection experiments.  The latter are optimized to detect
\textit{keV-scale} energy depositions from the elastic scattering of
WIMPs and the MeV-scale energy range is usually not reported.
Moreover, albeit similar detection principles and background rates,
$0\nu\beta\beta$ experiments are more readily scaleable in mass, and
the next decade is bound to produce progress not only in
limiting/detecting the Majorana mass for neutrinos,
but also in constraining the charged
excitations of WIMPs.
In what follows we consider representative models for both cases and
use them to set constraints from new data released this year by
EXO-200 \cite{Auger:2012ar} and Kamland-Zen \cite{KamLANDZen:2012aa}
$0\nu\beta\beta$ collaborations.

\paragraph{Representative Models}
\label{sec:models}

Case~A demands that the spins in the DM multiplet differ. With an eye
on SUSY we take the DM particle to be a Majorana fermion $X^{0}= \chi$
(like the neutralino) and the charged excited state to be a complex
scalar $X^{-} = \tilde\tau^{-} \equiv \tilde\tau$ (like the stau.) The effective
Lagrangian which governs the bound state formation is of Yukawa type
\begin{eqnarray}\label{differentspin} {\cal L}_{\rm A} &=& \bar\chi
  (g_{eL}\mathbb{P}_L + g_{eR} \mathbb{P}_R) e \tilde\tau^\dagger +
  {\rm h.c.} ,
\end{eqnarray}
with in general complex couplings $g_{eL,R}$ for the chirality
projections $ \mathbb{P}_{L,R}$.

In Case~B the DM states share the same spin. For simplicity, we take
both as scalars in the electroweak doublet, a real scalar
$X^0=\tilde\nu^{0}$ (like a sneutrino) and a complex scalar $X^{-} =
\tilde\tau^{-}$ as before. The relevant effective Lagrangian can be
written as
\begin{eqnarray}\label{samespin}
  {\cal L}_{\rm B} &=& \frac{g_{\rm eff}}{2}W^{-\mu} 
  (\partial_{\mu} \tilde\tau^{\dagger} \tilde\nu^{0} - 
  \tilde\tau^{\dagger}\partial_{\mu} \tilde\nu^{0}) + \rm{h.c.} + {\cal L}_{(4)}
\end{eqnarray}
where $ {\cal L}_{\mathrm{(4)}}$ contains quartic interactions. Assuming that 
the sneutrino state has small admixture of sterile states, in SUSY
$g_{\rm eff} = g_{2}\cos\theta_{\tilde\tau}$, %
where $g_2$ is the gauge coupling of the weak interaction and
$\theta_{\tilde\tau}$ is the mixing angle between the left and right
handed stau.

\paragraph{Bound state properties and formation}
\label{sec:bound-state-form}

\begin{table}[tb]
\caption{Relevant nuclei,  binding energies, and exposures~$MT$ for different experiments considered in this work.}
\begin{center}
\begin{ruledtabular}
\begin{tabular}{rccccccc}
Exps. ~&~ \begin{tabular}{c} 
EXO-200/ \\ Kaml.-Zen/ \\ Xe100 \end{tabular} ~&~\begin{tabular}{c} DAMA \\ NaI(Tl)\\ \end{tabular} ~&~ \begin{tabular}{c} SNO \\ NaCl\\ \end{tabular} & ~ \begin{tabular}{c} Bore-\\xino \end{tabular}   \\
\hline
 Nucleus ~&~ Xe ~&~ I ~&~ Cl ~&~ C \\
 $E_b^{(0)}$ (MeV) ~&~ 18.4  ~&~ 18.2 ~&~ 6.3 ~&~ 2.7 \\
$MT$ (kg yr)  ~&~ 40/30/0.9  ~&~ 7.5  ~&~ 1274 ~&~ $1.3\times 10^{5}$\\
\end{tabular}
\end{ruledtabular}
\end{center}
\label{table:exps}
\end{table}%

In Tab.~\ref{table:exps}, we list the ground state energies $E_b^{(0)}$ 
for the $X^-$ bound states with relevant nuclei, using the  homogeneous nuclear charge distribution 
within sphere of radius $R_0 = \sqrt{5  / 3}R_{\mathrm{rms}}$, and
$R_{\mathrm{rms}}$ values are taken
from Ref.~\cite{2004ADNDT..87..185A}.
Kinetic energies
$O(100\,\keV)$ of the incoming DM and the recoiling
bound state can be neglected, and the expected total visible energy $E_{\rm tot}$ injected in the
detector is given by
\begin{equation}
\label{eq:Etot}
E_{\rm tot} \approx 
\begin{cases}
E^{(0)}_{b} - \Delta m + m_e & \text{Case A, } \\
E^{(0)}_{b} - \Delta m + m_Z - m_{Z+1}    & \text{Case B.}
\end{cases} 
\end{equation}
In Case~A the positron is emitted with an energy $E_e^{(n,l)} =
E^{(n,l)}_{b} - \Delta m - m_e$, where $m_e$ is the electron mass and
$n, l$ denote the usual initial principal and orbital quantum numbers
of the capture.  After the positron is stopped, its annihilation with
an electron yields an additional energy of $2m_e$ whereas the excited
$(N\tilde\tau)^{*}$ relaxes by emission of gamma-rays. In case~B no
positron is emitted but the difference in nuclear mass upon the
nuclear transmutation $N_Z\to N_{Z+1}$ becomes accessible.

For case A, the product of recombination cross section~$\sigma_{A}$
with the incoming DM velocity $v$ reads,
\begin{equation}\label{sigmav}
\textstyle
  \sigma_{A} v \simeq  (|g_{eL}|^2 + |g_{eR}|^2)/(8\pi m_{\chi}) \times
  \sum_{n,l} B_{n,l}  ,
\end{equation}
where $B_{n,l} $ is the contribution from capture into state
$(n,l)$. In the limit  $\Delta m \gg m_e$ and $p_{\chi} \gg p_{e^{+}}$,
\begin{align}
  B_{n,l}& \simeq \left( E_{b}^{(n,l)} - \Delta m - m_e\right) \sqrt{( E_{b}^{(n,l)}  - \Delta m)^2 - m_e^2} \nonumber \\
  & \times \int d^3r_1d^3r_2\, \phi_{n,l}^*(\vec r_1)\phi_{n,l}(\vec r_2) e^{i
    \mu \vec v \cdot (\vec r_1 - \vec r_2)} .
\end{align}
Here, $\phi_{n,l}$ is the wave function of the relative motion of
$(N\tilde \tau)$ with reduced mass $\mu$. For $n\leq 50$ we calculate
$B_{n,l}$ explicitly by numerical solution of the Schr\"odinger
equation.
Whenever $\Delta m $ is small, capture into a multitude of states
(also with $n>50$) is possible. Importantly, in this case we expect
$\sigma_A$ to flow towards a semi-classical (SC) limit: when $\chi$
approaches the nucleus, a critical radius $r_b$ in the electromagnetic
potential $V(r)$ will be reached when the transition
$\chi\to\tilde\tau^{-} + e^{+} $ becomes energetically possible,
$V(r_b)+\Delta m + m_e \leq 0$. The integration over the fly-by time
when $r<r_b$ then gives the rate for this transition to happen.  In
this SC limit we find for $\sum_{n,l} B_{n,l}$,
\begin{align}
\nonumber 
 \int_{|\vec r|<r_{b}} 
\!\!\!\!\!d^3\vec r \sqrt{(V(|\vec r|)+\Delta m)^2 - m_e^2} ( - V(|\vec r|) - \Delta m - m_e) .
\end{align}
Whenever, say, $n>10$ is accessible in the capture, we find a perfect
agreement between the SC calculation and the explicit quantum mechanical calculation.
For lighter nuclei with shallow binding the latter is the preferred
choice as typically only few states contribute to the signal above the
detector threshold.  As expected, being an inelastic process,
$\sigma_A v$ is largely independent of $v$, {\em i.e.} 
$\langle \sigma_{A} v \rangle \simeq
\sigma_{A} v$.

Once $\Delta m $ and DM mass $m_{\chi}$ are chosen, the induced signal
in the neutrino searches can be translated into a constraint on the
combination of Yukawa couplings $(|g_{eL}|^2 + |g_{eR}|^2)$. We choose
to trade the latter against a constraint on the $\tilde\tau$ lifetime
instead, or, equivalently on its decay width $\Gamma_{\tilde\tau} =
\tau_{\tilde\tau}^{-1}$. In the same approximations,
\begin{eqnarray}
  \Gamma_{\tilde\tau} &\simeq& \frac{\sqrt{\Delta m^2 - m_e^2} }{4\pi m_\chi} 
(\Delta m+m_e)(|g_{eL}|^2 + |g_{eR}|^2) .
\end{eqnarray}

Let us now consider case~B. During the capture an interconversion from
a neutron $(n)$ to a proton $(p)$ takes place inside the
nucleus. We use the Fermi gas model for the density and momentum distributions
of $n$ and $p$ inside the nucleus $N_Z$. The calculation of $\sigma_B$
proceeds in two steps. First, we compute the fundamental process of
$n\to p$ in the presence of $\tilde\nu^0$.  Because of Pauli blocking,
part of the binding energy is invested in elevating $p$ above the
Fermi surface of $N_{Z+1}$.
In a second step we again obtain the total cross section by
consideration of the fly-by time of $\tilde\nu^0$ in which the
transition is possible. The result reads,
\begin{align}
  \sigma_B v  &= \frac{g_{\rm eff}^4 m_p^2}{8 M_W^4} \int d^{3}r \rho_n(\vec r) \int^{m_n+\frac{p_{nF}^2}{2 m_n}}_{m_{n}}  \frac{d p_{n}^{0}}{\frac{4}{3}\pi p_{pf}^3} \nonumber\\
  & \times {\sqrt{(-V(r)-\Delta m+p_{n}^{0})^{2}- m_{p}^{2}} \sqrt{{p_{n}^{0}}^{2} - m_{n}^{2}}} \nonumber\\
  & \times \theta \left(- V(r) - \Delta m + p_{n}^{0} - m_{p} - \frac{p_{pF}^{2}(N_{Z+1})}{2 m _{p}}\right) , \nonumber\\
\end{align}
where $\rho_n$ is the number density of neutron inside the target
nucleus and $p_{pF}(N_{Z+1})$ is the Fermi momentum of the $N_{Z+1}$
nucleus. The latter is determined from $m_p + \frac{p_{pF}^2(N_{Z+1}
  )}{2 m_p} = m_n + \frac{p_{nF}^2( N_{Z} )}{2 m_n}$. For the
capture on xenon, $  N_{Z} = \xe  $ and  $  N_{Z+1} = \cs  $. 

This calculation is SC, and is valid for small values of $\Delta m$,
when the energy release is significant relative to the typical nuclear
level intervals, and many bound states are available for capture.
Without detailed knowledge of the nuclear wave functions, an exact quantum mechanical
calculation is not possible, although it is expected that the cross
section can be dominated by a series of narrow resonances especially
when $\Delta m$ is large \cite{Pospelov:2008qx,Bai:2009cd}.

\paragraph{Neutrino experiments}
\label{sec:neutr-double-beta}

We now come to the central part of our study---demonstrating the
potential of neutrino experiments and in particular of
$0\nu\beta\beta$ searches in probing DM. As we have argued, the total
energy in (\ref{eq:Etot}) is monochromatic with all its energy
injected almost instantaneously. The signal shape is hence determined
by the energy resolution $\sigma_e$ of the experiment. Given an
exposure $MT$ the total number of events in an energy bin $\Delta E =
E_{\mathrm{max}} - E_{\mathrm{min}}$ then reads,
\begin{align}
  N_{\rm tot}& = \frac{MT N_T  \rho_{\rm DM} \langle\sigma
    v\rangle}{ 2m_{X^0}} \nonumber \\
& \!\!\!\!\!\!\! \times  \left[ \erf \left(
      \frac{ E_{\mathrm{max}} - E_{\mathrm{tot}}  }{\sqrt{2}\sigma_e}\right)
    - \erf \left(
      \frac{ E_{\mathrm{min}} -  E_{\mathrm{tot}}   }{\sqrt{2}\sigma_e}\right)
  \right] ,
\end{align}
where $\rho_{\rm DM}\approx 0.3~{\rm GeV/cm^3}$ is the local DM energy
density, $N_T$ is the number of target nuclei per kg of  active detector
mass.

\begin{figure*}[tp]
\begin{center}
\includegraphics[width=0.41\textwidth]{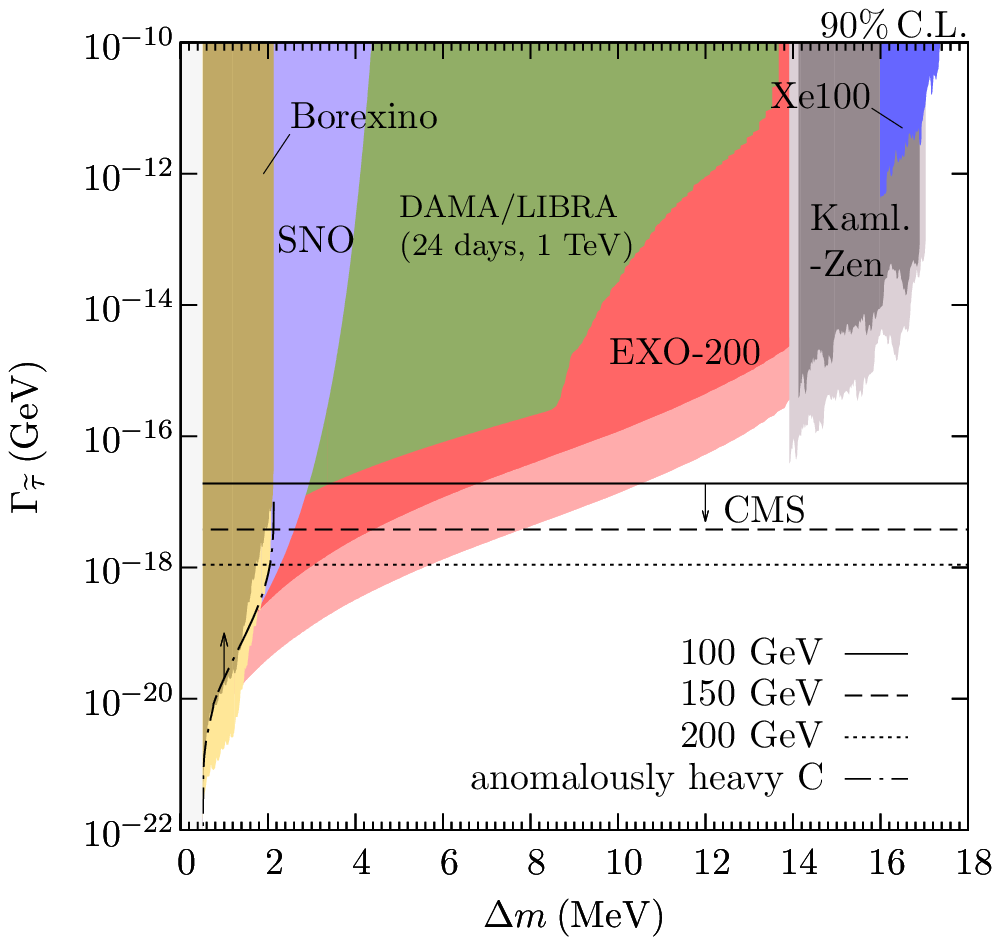}%
\hspace*{1cm}
\includegraphics[width=0.41\textwidth]{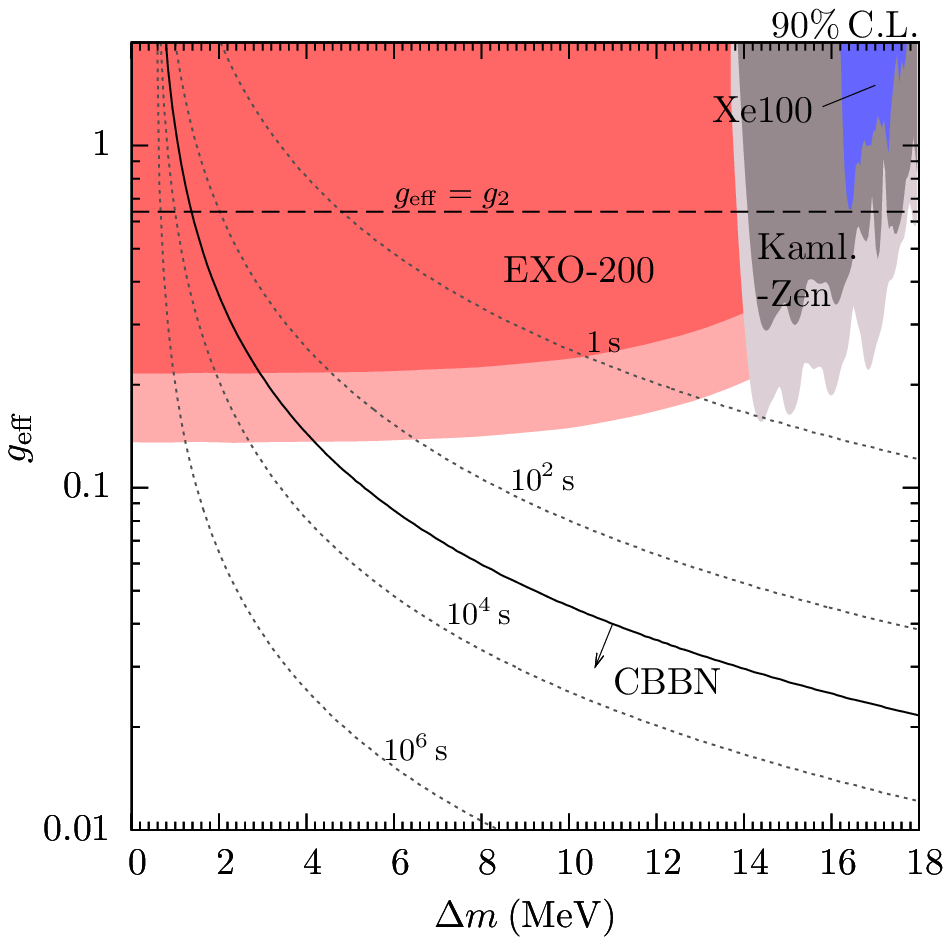}
\caption{\textit{Left:} Constraints for ``Case~A'' from underground
  rare event searches as labeled. Darker (lighter) shadings are for
  $m_{\chi}=1\,\TeV$ (100\,\GeV). Also shown are lower limits from CMS
  searches on long-lived charged states. \textit{Right:} Corresponding
  constraints for ``Case~B''. The dotted lines are contours of
  constant $\tilde\tau$ lifetime; lighter shadings are for
  $m_{\chi}=200\,\GeV$. }
\label{fig:combined}
\end{center}
\end{figure*}

We start by considering the EXO-200 experiment~\cite{Auger:2012ar}
which has 110~kg fiducial mass enriched in \xex, whereas for our purpose the entire mass of Xe in the fiducial volume
becomes active, and the exposure as a DM target is listed in Tab.~\ref{table:exps}. The TPC-type setup of EXO-200
allows to distinguish between events occurring at a single sites (SS)
or at multiple sites (MS). The capture typically qualifies as an MS
event, given the macroscopic range of produced $\gamma$-rays
($\approx6~\cm$ at 1~MeV in LX.) We use the last bin containing 24
events above 3.5~\MeV\ to set a constraint which thereby probes the
region $\Delta m \lesssim 15~\MeV$.
Two complications arise: 1) part of $E_{\mathrm{tot}}$ can be taken
out of the fiducial volume by $\gamma$-rays and 2) events at SS with
depositions greater than 10 MeV are automatically discarded. We have
studied both effects in a dedicated Monte Carlo (MC) analysis (to be presented in detail elsewhere) 
and find
that this can weaken the limits by up to a factor of two, and which is
taken into account in presented constraints. 
The results for cases A and B are shown in Fig.~\ref{fig:combined}

The next experiment we consider is Kamland-Zen
\cite{KamLANDZen:2012aa}.
To obtain a conservative constraint, we use
the reported spectrum from which we only subtract the background from
the $2\beta$ decay of $\xex$. The energy resolution is excellent so
that we only need to consider the bin in which $E_{\mathrm{tot}}$ lies
including the neighboring ones (with appropriate statistical penalty)
in deriving the constraint.
Unfortunately, Kamland-Zen does not report a \textit{cumulative} last
bin like EXO-200 does. Therefore, we can only probe energy depositions
below the last reported bin at $3.8~\MeV$. This, in turn, restricts
the considered mass splittings to $\Delta m \gtrsim 14\,\MeV$ as shown in Fig.~\ref{fig:combined}.  
Once $\Delta m $ becomes large enough to approach the threshold, the
limits become correspondingly weaker because of limited phase space
and higher count rates.

Turning to small values $\Delta m \lesssim 4\,\MeV$ we either probe
energy depositions beyond 10~MeV with heavy elements like with $\xe$
or, alternatively, we can probe smaller (and hence reported) energy
depositions using lighter nuclei. 
For the latter, the neutrino experiments of choice are currently the
salt phase of the SNO water Cherenkov detector with a 2~ton loading of
NaCl and the Borexino detector with its carbon-based scintillator.  We
set constraints using the respective event spectra reported
in~\cite{Aharmim:2005gt} and~\cite{Bellini:2012kz} where in the former
case we subtract the solar neutrino signal. The excluded regions are
shown on-top of the EXO-200 one. 
Finally, we note that $\Delta m \lesssim 4\,\MeV$ is also challenged
by severe limits~\cite{Hemmick:1989ns} of anomalously heavy C, N, O
nuclei among others. From Fig.~\ref{fig:combined} it can be seen that
the limit from Borexino is comparable.

\paragraph{Direct DM detection experiments}
\label{sec:direct-detect-exper}

One of the current most sensitive direct detection experiments is
XENON100. Since it is a LX detector, we can directly compare it to
EXO-200 and Kamland-Zen. Using the electromagnetic background spectrum
up to 3~MeV published from a commissioning run~\cite{Aprile:2011vb},
we find that constraint is at least one order of magnitude
weaker. This can be traced back to higher levels of radioactive
background present in XENON100.

A competitive experiment in sensitivity can be DAMA/LIBRA with its
250~kg of NaI scintillating crystals. A high-energy ``single-hit''
spectrum from 9 out of 25 crystals of a 24~day run was reported
in~\cite{Bernabei:2009zzb}. No events above 10~MeV were observed.  A
conservative constraint in case A is obtained by considering only
recombination into the 1$S$ ground state of I because no rays are produced. The latter may well escape one crystal challenging a
single-hit observation.  For this reason we refrain from studying
case~B which certainly requires MC modeling. For case~A we find that
the constraint is superseded by the one of EXO-200.  We stress that if
DAMA were to present us with a high-energy spectrum of a larger
exposure, a very strong result could be obtained.

\paragraph{Constraints from Colliders}
\label{sec:constr-from-coll}

For heavy charged stable particles, the CMS group placed a very
stringent bound based on the data with an integrated luminosity of
5~fb$^{-1}$ and a centre-of-mass energy of
7~TeV~\cite{Chatrchyan:2012sp}. 
For case~A we obtain a limit on $\Gamma_{\tilde\tau}$ from the
constraint on the LHC $\tilde\tau$ electroweak production cross section
$\sigma_{\rm production} $,
\begin{equation}
\sigma_{\rm production} \times \exp \left(  - R\Gamma_{\tilde\tau}/ \gamma v_T \right)  < \sigma_{\rm constraint} \ ,
\end{equation}
where $R$ is the transverse distance, $v_T$ the transverse velocity
and $\gamma$ the boost factor; $\sigma_{\rm constraint}$ we take
from~\cite{Chatrchyan:2012sp}. The lower limits on
$\Gamma_{\tilde\tau}$ for $m_{\chi} = 100$~GeV (and above) are shown
in the left Fig.~\ref{fig:combined}. The combined constraints rule out
the entire region $\Delta m < 5$~MeV as long as $m_{\chi} < 210$~GeV;
larger values of $m_{\chi}$ are currently not constrained by CMS. In the case that the excited state is a spinor, the constraint will become stronger because the dominant production channel is s-wave, instead of p-wave in the scalar case. 
In case~B, $\tilde\tau$ escapes the detector because
$\Gamma_{\tilde\tau}$ is suppressed by $\Delta m^{5}/m_{W}^4$. The CMS
constraint is then: $m_{\tilde\nu} > $ 210~GeV. Closely related searches
are also performed by ATLAS, but 
specialized to specific SUSY cases, and we do not use it here.

Finally, for lifetimes $\tau_{\tilde\tau} \gtrsim 10^3$~s a primordial
abundance of $\tilde\tau^{-}$ (which is expected to be similar to that
of $\tilde\nu^0$ since $\Delta m/m_{\tilde\nu}\ll 1\%$) will have a
large effect on the primordial light elements;
see~\cite{Pospelov:2010hj} and references therein. The approximate
constraint is labeled~``CBBN''.

\paragraph{Outlook}
\label{sec:discussion-outlook}

We have shown that rare underground event searches can be used to
obtain strong limits on charged excited states of DM with mass
splittings $\Delta m\lesssim 20~\MeV$. Significant improvements on
these bounds can be expected from nearly all upcoming $0\nu\beta\beta$
searches, %
see \cite{Elliott:2012sp} and references therein, as well as
from future data sets of Kamland-Zen and EXO-200. 
In some cases great sensitivity is already experimentally established
but %
not published. For example, Kamland-Zen has not
reported its count rate above 3.8~\MeV. Presumably, the associated
constraint can be the dominant one throughout the entire considered
region in $\Delta m$. Likewise, DAMA has shown a fraction of its high
energy spectrum with an exposure of 24 days---whereas more than 100
times of this is available.
We urge these
collaborations to present this data. 
\newline
{\em Acknowledgements} We thank Drs.~J.~Chen, M.~Marino, A.~Ritz, ~N.~Toro for useful discussions. 

\end{document}